# Potemkin village of Russian science: The case of JINR

Tetiana Hryn'ova (LAPP/IN2P3/CNRS)

**Abstract**

**This document reviews international cooperation of the Joint Institute of Nuclear Research (JINR). It shows that of the original 11 founding members of the JINR, only one - Mongolia - is still fully active within the organization. Out of 16 existing members four have their participation suspended. The review of the Topical Plan for JINR Research and International Cooperation 2024 shows that the number of 990 institutes collaborating with JINR is overestimated by at least 10% and probably closer to 25%. Countries with the highest number of participating institutes listed for 2024 are Russia, USA and Italy with 203, 77 and 49 institutes, respectively. JINR collaboration with impostor institutes on the occupied territories of Ukraine is presented as well as JINR publications which associate the occupied Ukrainian territories to Russia. JINR cooperation with organizations in Russia, which are under sanctions for their support of Russia's invasion of Ukraine and contribution to the war effort, is also discussed.**

**Section 1. Introduction**

This report is the second in a series of "Potemkin village of Russian science", which studies the response of the international academic community after more than two years of Russia's full-scale invasion of Ukraine.

The first report [1] presented results of a survey of international members of editorial boards of English-language journals edited by the institutes in the Russian Federation. It was found that 70.9% of them were either unaware of their membership, or did not fulfill their editorial duties, or were retired, or deceased. Such deception is not only used to enhance the profile of those journals by also useful for propaganda inside Russia to argue that these scientists and by extension the international scientific community approves actions of the associated Russian institutions, many of which are under sanctions for their support of Russia's invasion of Ukraine and contribution to the war effort.

This report focuses on the Joint Institute for Nuclear Research (JINR), a scientific center located in Dubna, Russia and mainly funded by it. This report relies on information from the JINR website [2], in particular its Topical Plans for JINR Research and International Cooperation (TP), which are released yearly in English [3]. Notably the Russian-language version of the topical plans [4] has additional information on the scientific contacts at collaborating institutes.

Section 2 overviews status of JINR member and associate member states. Section 3 discusses the number of organizations collaborating with JINR and their distribution throughout the world. Section 4 presents organizations under sanctions collaborating with JINR. Section 5 presents the JINR position with respect to temporary occupied territories of Ukraine. Section 6 discusses their association to the Russian Federation in the journals edited by JINR. Section 7 overviews membership in the JINR advisory bodies. Conclusions are presented in Section 8.

**Section 2. JINR member and associate member states**

JINR was created 1956 as an international intergovernmental scientific research organization by 11 founding states: Albania, Bulgaria, Hungary, German Democratic Republic, Republic of China, Korean Democratic Republic, Mongolia, Poland, Romania, USSR[1], and Czechoslovakia.

In 2024, JINR has sixteen member states: Armenia, Azerbaijan, Belarus, Bulgaria, Cuba, Arab Republic of Egypt, Georgia, Kazakhstan, D. P. Republic of Korea, Moldova, Mongolia, Romania, Russia[2], Slovakia, Uzbekistan, and Vietnam. Participation of the D. P. Republic of Korea was suspended in 2015 [5]. Topical plans for 2023 and 2024 contain a note for Bulgaria, Romania and Slovakia that "The cooperation may be limited by the conditions adopted unilaterally by the State". This is due to suspension of their JINR membership [6]. Thus, of 16 current JINR members, four have suspended their participation. As only Bulgaria, D. P. Republic of Korea, Mongolia and Romania remain from the original list of the founding countries, this leaves only one founding state - Mongolia - still fully active within JINR.

JINR also claims to have five associate member states: Germany, Hungary, Italy, the Republic of South Africa and Serbia. Their status is less well-defined and associate state membership contacts given on JINR website only list the country representatives for Serbia. Topical plans for 2023 and 2024 for the case of Germany also state that "The cooperation may be limited by the conditions adopted unilaterally by the State", which could indicate suspension of the common research activities by Germany.

**Section 3. Number of organizations collaborating with JINR**

As shown in Fig. 1, on its website JINR claims collaboration with 990 research centers and universities. Topical plan for 2024 was reviewed below to understand the origin of this number.

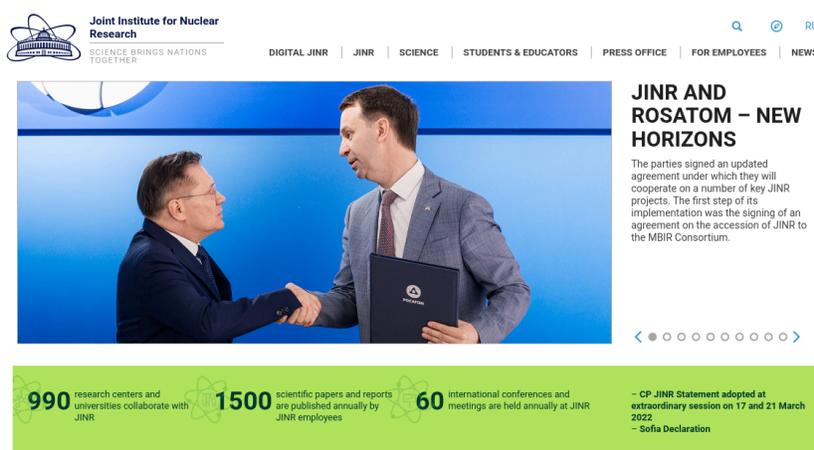

**Fig. 1** From JINR website http://www.jinr.ru/about-en/. Image taken on Oct 19, 2024.

---

[1] The USSR contributed 47.25% of the JINR budget.
[2] Russia contributed over 80% of the budget as of 2022.

**Section 3.1 Overview of organizations collaborating with JINR from JINR member states**

Number of JINR collaborating institutes per JINR member state is shown in Table 1. It is notable that Romania and Bulgaria, second and third countries in this list, limited or suspended their participation in the JINR activities. The total number of institutes in this category is 366. It decreases to 306 if Romania, Bulgaria and Slovakia are removed from the calculation. Note that over 55% of organizations in this category are based in Russia, which becomes over 66% if organizations from Romania, Bulgaria and Slovakia are removed from the calculation.

**Table 1.** Number of organizations collaborating with JINR per JINR member state country. Asterix means that for that country "The cooperation may be limited by the conditions adopted unilaterally by the State".

| Country | N institutes | Country | N institutes |
| --- | --- | --- | --- |
| Russia | 203 | Kazakhstan | 9 |
| Romania* | *33* | Slovakia* | *9* |
| Bulgaria* | *18* | Azerbaijan | 8 |
| Belarus | 17 | Moldova | 7 |
| Armedia | 11 | Mongolia | 7 |
| Egypt | 11 | Cuba | 6 |
| Uzbekistan | 11 | Georgia | 6 |
| Vietnam | 10 | N. Korea | 0 |

**Section 3.2 Overview of organizations collaborating with JINR from JINR associate member states**

**Table 2.** Number of organizations collaborating with JINR per JINR associate member state country. Asterix means that for that country "The cooperation may be limited by the conditions adopted unilaterally by the State".

| Country | Number of Institutes |
| --- | --- |
| Italy | 49 |
| Germany* | *37* |
| South Africa | 14 |
| Serbia | 10 |
| Hungary | 5 |

4Number of JINR collaborating institutes per JINR associate member state is shown in Table 2. It is notable that Germany, the second country on this list, limited or suspended its participation in the JINR activities. The total number of institutes in this category is 115. It decreases to 78 if Germany is removed from the calculation.

**Section 3.3 Overview of organizations collaborating with JINR from other states beyond Europe and Asia**

Number of JINR collaborating institutes for other countries outside Europe and Asia is presented in Table 3. There are three countries participating from North America, with a total of 87 institutes. This list is heavily dominated by the USA, which has the highest number of institutes listed as collaborating with JINR after Russia. South America, Africa and Oceania have 4, 2 and 2 countries listed each with 18, 5 and 2 institutes, respectively.

**Table 3.** Number of organizations collaborating with JINR for non-member countries outside Europe and Asia.

| Continent/Country | Numb. of Inst. | Continent/Country | Numb. of Inst. |
| --- | --- | --- | --- |
| **North America** | **87** | **South America** | **18** |
| USA | 77 | Brazil | 13 |
| Mexico | 6 | Chile | 3 |
| Canada | 4 | Argentina | 1 |
|  |  | Peru | 1 |
| **Oceania** | **5** | **Africa** | **2** |
| Australia | 3 | Botswana | 1 |
| New Zealand | 2 | Tunisia | 1 |

**Section 3.4 Overview of organizations collaborating with JINR from other states in Europe**

Number of JINR collaborating organizations for other countries in Europe is presented in Table 4. There are 27 participating countries listed in the Topical plan for 2024, although it is known that Poland, Czechia and Ukraine officially stopped their participation in the JINR activities in 2022 and thus their institutes should not be listed. Also, a number of listed institutes in France is questionable: many of them are either duplicates or seized to exist a long time ago as is discussed in Appendix A.

In any case the total number of other organizations collaborating with JINR in Europe is at most 167. There are 42 institutes listed for Poland, Czechia and Ukraine and at least 8 institutes are non-existing or duplicates for France. Removing those leaves 117 institutes in this category.

**Table 4.** Number of organizations collaborating with JINR for non-member countries in Europe.

| Country | N of Inst. | Country | N of Inst. | Country | N of Inst. |
|---|---|---|---|---|---|
| France | 35 | Sweden | 5 | Albania | 1 |
| *Poland** | *19* | Switzerland | 5 | Cyprus | 1 |
| UK | 18 | Croatia | 4 | Denmark | 1 |
| *Czechia** | *16* | Netherlands | 4 | Estonia | 1 |
| Spain | 14 | Norway | 4 | Latvia | 1 |
| Belgium | 8 | Portugal | 3 | Montenegro | 1 |
| *Ukraine** | *7* | Austria | 2 | Malta | 1 |
| Greece | 5 | Ireland | 2 | N Macedonia | 1 |
| Finland | 5 | Lithuania | 2 | Slovenia | 1 |

**Section 3.7 Overview of organizations collaborating with JINR from other states in Asia**

Number of JINR collaborating organizations for other countries in Asia is presented in Table 5. There are 14 participating countries listed in the Topical plan for 2024, with 121 institutes listed. The top participating countries are India, Japan, China and South Korea.

**Table 5.** Number of organizations collaborating with JINR for non-member countries in Asia.

| Country | N of Inst. | Country | N of Inst. |
|---|---|---|---|
| India | 25 | Israel | 3 |
| Japan | 25 | Pakistan | 3 |
| China | 22 | Taiwan | 3 |
| South Korea | 18 | Tajikistan | 3 |
| Turkey | 7 | Bangladesh | 1 |
| Thailand | 5 | Indonesia | 1 |
| Iran | 4 | Sri Lanka | 1 |

**Section 3.8 International organizations collaborating with JINR**

There are also two international organizations International Atomic Energy Agency (IAEA) in Vienna and European Organization of Nuclear Research (CERN) in Geneva, which are listed among JINR collaborating institutes. After 2022, JINR participation at CERN is affected by a number of the CERN Council Resolutions [8].

**Section 3.9 Discussion on the overall number of organizations collaborating with JINR**

The current study did not investigate systematically the validity of individual projects and claimed collaborations of JINR with organizations in other countries. But when approached for information, a few contacts listed in the Russian version of the Topical Plan of 2024
- were not aware of their participation in the project listed or any collaboration between JINR and their institutes or
- were not associated with institute listed in the JINR TP2024, or
- were no longer in charge of the project listed or
- were retired or dead[3].

Thus understanding the real situation as for international collaboration status of JINR requires dedicated and systematic study, similar to that performed in [1].

As shown in Table 6, the total number of organizations collaborating with JINR as listed in its Topical Plans for 2024 is at most 883 and is probably significantly lower than even the corrected estimate of 726. This is 10% and 25% lower than the number of 990 listed on the JINR website as is shown in **Figure 2**.

**Table 6.** Summary table of countries and organizations collaborating with JINR as per TP2024. Number in brackets corresponds to the countries marked with * in the TP2024 and North Korea.

| Category | Countries | Institutes in TP2024 | Institutes Corrected |
|---|---|---|---|
| Member States | 16 (4) | 366 | 306 |
| Associate Members | 5 (1) | 115 | 78 |
| North America Other | 3 | 87 | 87 |
| South America Other | 4 | 18 | 18 |
| Oceania Other | 2 | 5 | 5 |
| Africa Other | 2 | 2 | 2 |
| Europe | 27 (3) | 167 | 107 |
| Asia | 14 | 121 | 121 |
| Total Countries | 73 | 881 | 724 |
| Organizations | 2 | | |

---

[3] Dead scientists listed as contacts for Ukrainian institutes G. Zinoviev (died in 2021, p 27, 129), K. Bugaev (died in 2021, p 27), I. Zalyubosvski (died in 2013, p 27). Page numbers from the Russian version of TP 2024.

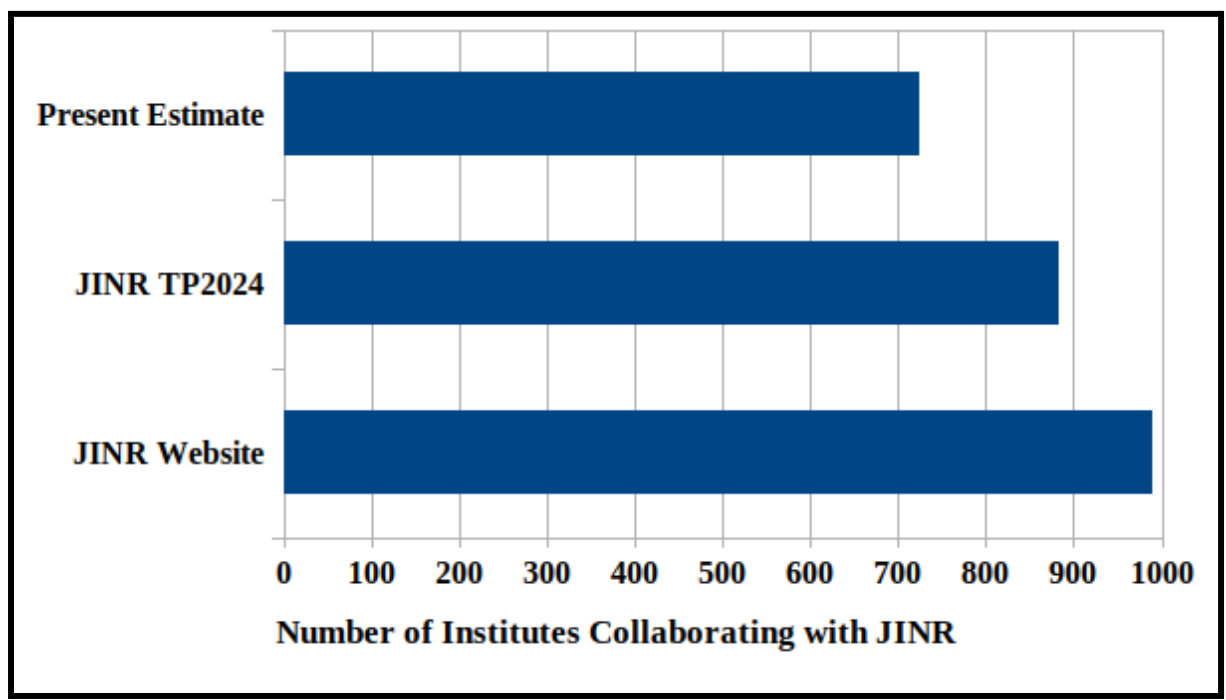

**Figure 2.** Number of Institutes Collaborating with JINR from various sources. Present estimate value could still be an overestimation of the real number.

It is interesting to note that JINR removed institutes from N. Korea from its Topical Plans list, but has not done the same for other countries which terminated or suspended its activities with it. There is also a high number of organizations which are included on the list of JINR collaborators only through a single common project in a big international experiment (e.g. ATLAS, CMS or ALICE at CERN or DUNE at Fermilab). For example, all institutes in Austria, Bangladesh, Cyprus, Denmark, Indonesia, Netherlands, Malta, Montenegro, Sri Lanka, 4 out of 5 institutes in Thailand fall into this category.

**Section 4. Organizations under sanctions collaborating with JINR**

> JINR is a city-forming enterprise of the science city of Dubna. The Institute's active participation in formation and development of the Dubna innovation ecosystem, partnership with the SEZ "Dubna" and its residents open up wide opportunities for technology and knowledge transfer.

**Figure 3.** From the JINR website [https://www.jinr.ru/innovations-en/], captured on 20/10/2024.

As shown in **Figure 3**, JINR was fundamental for creation of the Special Economic Zone (SEZ) of Dubna in 2005, which hosts multiple enterprises involved in Russian military projects and which are currently under sanctions [7]. JINR itself and scientists and engineers working there created multiple enterprises in SEZ.

JINR was also fundamental for the creation of Dubna State University (DSU) in 1995. DSU trains personnel for the enterprises working on the weapon's development in the SEZ Dubna [7]. It is listed in the OpenSanctions database as a sanction-linked entity. JINR scientists teach in DSU, while DSU professors and students participate in the JINR research projects.

Both SEZ Dubna and DSU are listed as JINR collaborating institutes in the Topical Plan 2024.

There are also 78 JINR collaborating organizations from Russia which are subject to sanctions due to their contributions to Russia's weapon development and support of the war in Ukraine: 19 in EU, 34 in USA 34, 19 in Switzerland, 7 in the UK, and 70 in Ukraine (see list in Appendix B for the full list).

JINR itself is listed in the OpenSanctions database as a sanction-linked entity, as shown in **Figure 4.**

**Section 5. JINR position with respect to the temporary occupied territories of Ukraine**

Since 2014, all Ukrainian institutes have relocated from territories occupied by Russia to other regions of Ukraine. However, Russia has created impostor entities using the stolen assets of Ukrainian scientific organizations and appropriating their names. **Table 7** (extracted from Ref. [11]) shows a selective list of such entities, which JINR collaborated with and claimed as being located in Russia.

JINR regularly holds international conferences and meetings in the occupied territories, primarily at its "Dubna" resort in the occupied Alushta, which it claims to be located in Russia. Examples of such events are yearly International Scientific Workshop in Memory of Professor V.P. Sarantsev "Problems of Colliders and Charged Particle Accelerators" or 27th Russian particle accelerator conference in 2021.

**Section 6. JINR journals**

JINR has two English-language journals, distributed by the international publisher Springer Nature: Physics of Particles and Nuclei and Physics of Particles and Nuclei Letters. The Editorial Boards of these JINR-run journals are not immune to issues discussed in Ref. [1].

These journals also host publications and conference materials where occupied territories of Ukraine are presented as Russia [9, 10], notably the proceedings for 2017 International Scientific Workshop in Memory of Professor V.P. Sarantsev "Problems of Colliders and Charged Particle Accelerators", which were published by Springer in the JINR-run journal Physics of Particles and Nuclei Letters Volume 15, Issue 7 (2018).

These publications are disseminated on international scientific websites, in journals, and in databases [11,12], supporting the Russian government's goals of legitimizing and normalizing the Russian occupation of Ukrainian territories within the scientific community.

# Joint Institute for Nuclear Research

**Sanction-linked entity**

Joint Institute for Nuclear Research is an entity of interest. It has not been found on international sanctions lists.

| | | |
|---|---|---|
| **Type** | Organization | [sources] |
| **Name** | Joint Institute for Nuclear Research | [sources] |
| **Legal form** | not available | [sources] |
| **Country** | Russia | [sources] |
| **Description** | Research facility. According to a Member State, Kim Il Sung University, the Kim Chaek University of Technology, the Pyongsong College of Science and the National Defense University are the main institutions feeding the nuclear or ballistic missile programmes of the country. Has engaged in various academic cooperations, student and staff exchanges with international universities. "The Panel investigated possible technical exchanges between Kim Chaek University of Technology and International Global Systems (M) Sdn Bhd and International Golden Services Sdn Bhd in Kuala Lumpur, which appeared to have been front companies of Pan Systems Pte. Ltd. (also known as Glocom). Malaysia stated that these companies ceased operation in July 2011 and February 2014 and were dissolved in January 2019 and June 2018, respectively. It also explained that "Malaysian authorities have no information on any possible technical exchange between [Kim Chaek University of Technology] and [the companies], which is related to [Democratic People's Republic of Korea] nuclear programme", and confirmed that "there is no [Democratic People's Republic of Korea] national currently living/ working in Malaysia"." | [sources] |
| **Website** | jinr.ru | [sources] |
| **Status** | not available | [sources] |
| **Address** | Dubna | [sources] |
| **Last change** | 2024-06-18 **Last processed** 2024-10-09 **First seen** 2023-08-08 | |

## Relationships

**Members**

| Member | Role | Start date | End date | |
|---|---|---|---|---|
| Kim Son Hyok | Member of the Scientific Council | - | - | |
| Hong Mi Dok | Laboratory technician at the Laboratory of Nuclear Problems | - | 2015-03-30 | |
| Rim Yong Chol | Researcher, Laboratory of Information Technologies | 2012-04 | 2015-03-30 | |
| Pak Ben Seb | Former Scientific Council member | - | - | |
| Ri Je Son | Member of the Committee of Plenipotentiaries of the Governments of the 18 Member States | 1998 | - | |
| Ri Yong Suk | Member of the Laboratory of Nuclear Problems | - | 2015-03-30 | |
| HWANG SOK-HWA | Former Scientific Council member | 2009-07-16 | - | |

## Data sources

**DPRK Reports** — 5,540

A database of entities and events related to North Korea's sanctions evasion efforts.

United Kingdom · RUSI · non-official source

**Figure 4** OpenSanctions database entry on JINR. Captured on 20/Oct/2024.

**Table 7** Selected impostor organizations on the occupied territories of Ukraine and the occupied Ukrainian locations which JINR declares to be in Russia in its TP2024 or publications cited in "Refs." column. The approximate date of impostor institute creation is given in brackets, if known, together with its current sanctions status.

| Impostor Institutes created by Russia (different spellings exist) | Original Ukrainian Institutes | Changed location | Refs. | Locations |
|---|---|---|---|---|
| A. O. Kovalevsky Institute of Biology of the Southern Seas of RAS (2014), under sanctions | A. O. Kovalevsky Institute of Biology of the Southern Seas of NANU renamed in 2014 into Institute of Marine Biology of NANU, currently in Odesa | Y | TP2024 [13], p 222 | Sevastopol, Crimea |
| Donetsk State University (2014) | Vasyl' Stus Donetsk National University which was moved to Vinnytsia | Y | [9] | Donetsk |
| Donetsk Institute for Physics and Engineering named after A. A. Galkin (2016) | O. O. Galkin Donetsk Institute for Physics and Engineering of NANU, currently in Kyiv | Y | TP2024 [13], p 218 | Donetsk |
| Donetsk National Technical University (2014) | Donetsk National Technical University currently in Lutsk | Y | [9] | Donetsk |
| Crimean Astrophysical Observatory of RAS (2015) | Research Institute "Crimean Astrophysical Observatory", Ministry of Science of Ukraine | N | [10] | Nauchny, Katsiveli, Simeiz, Yalta, Crimea |

**Section 7. Membership on the JINR advisory bodies**

Distribution of scientists participating in the JINR scientific council (SC) and Program advisory committees (PAC) is shown in **Table 8**. Highest number of scientists (15) comes from the institutes in Russia[4]. The second highest number of participants, 6 members, comes from Bulgaria, despite its suspension of participation in the JINR activities. Of the associate member states, Italy has the highest representation rate with 3 members. Non-member states with notable SC and PACs participation are China (4 scientists), France (4) and India (4). There are also scientists from the USA (2.5[5]) and Mexico (1) in North America; South Korea (3) and Israel (2) in Asia; Latvia (1) and Switzerland (0.5) in Europe. There is also participation of scientists from other countries which limited their participation with JINR: such as Germany (1) and Romania (3). Overall nearly 30% of scientists in the Particle Physics PAC have chosen to put their membership on hold since 2022. They come from institutes in Poland, Czechia and Germany as well as CERN.

**Table 8** Affiliations of scientists participating in the JINR scientific council (SC) and Program advisory committees (PAC) as of October 23, 2024. The value in brackets corresponds to the number of scientists who suspended their membership.

|  | SC | PAC Particle Physics | PAC Nuclear Physics | PAC Condensed Matter |
|---|---|---|---|---|
| Russia | 7 | 2 | 3 | 3 |
| Member States (w/o Russia) | 19 | 2 | 1 | 4 |
| Associate Members | 3 | 2 (1) | 2 | 3 |
| Other in Europe | 2 | 2.5 (2) | 1 | 2 |
| Other in Asia | 5 | 3 | 4 | 1 |
| Other in North America | 2 | 2.5 | 0 | 0 |
| Other in South America | 2 | 1 | 1 | 0 |
| CERN | 0 | 2 (2) | 0 | 0 |
| Total | 40 | 17 (5) | 12 | 13 |

**Section 8. Conclusions**

Despite JINR collaborating with many organizations under sanctions, its scientists continue to partake in research projects throughout the world (even if potentially less significantly than presented in the JINR documents) and international scientists continue to participate in the JINR research activities, JINR advisory bodies and editorial boards of its journals.

---

[4] This number includes JINR.
[5] A fractional value is assigned to a scientist which has dual affiliation.

This document shows that of the original 11 founding members of the JINR, only one - Mongolia - is still active within the organization. Out of 16 existing members four have their participation suspended. The total number of collaborating institutions claimed to be 990 on the JINR website is shown to be overestimated by at least 109 institutes (10%), although that number should be more than doubled to 266 (over 25%). This includes continuing listing of institutes from the countries which terminated their JINR participation as well as duplicate institutes; scientists who are retired or dead or unaware of the projects they are listed as coordinating in the JINR Topical Plans for 2024.

Similarly to the situation described in Ref. [1], such deceptions are used to enhance the profile of JINR within the international scientific community to encourage continuation of existing collaborations as well as to fostering new ones. It is also used for propaganda purposes inside Russia to make JINR and its collaborating organizations in Russia more attractive to young scientists.

The questions posed most often in response to this study were: "What can be done?" and "How to ensure that JINR corrects the wrong information?". It is clearly not in the interest of the JINR management to correct the erroneous information (which would indicate the decrease of its standing within the international community). While it is important for the international scientific community to ensure the correct presentation of its relations with JINR, both in terms of participating institutes, projects and contributing scientists. The verification of these can be done by individual scientists, representatives of the scientific institutes or by funding agencies based on the same literature as the study presented in this report [13]. Notably, the JINR Scientific Council and Program Advisory Committees members still have scientists from most of the continents included. It is their scientific responsibility to ensure the correctness of the documents they review, such as the Topical Plan for JINR Research and International Cooperation.

**Appendix A Duplicate and non-existing French institutes in Topical Plan 2024**

Since 2019 LMS/Modane is attached to LPSC Grenoble (which is also present on the list) and thus can be removed as duplicate [https://phototheque.in2p3.fr/index.php?/category/1429].

CSNSM/IPN and LAL in Orsay united into IJClab in Orsay (which is already listed separately) in 2020, and thus can be removed as duplicates.

CRN/Strasbourg was dissolved in 1997 to become IRES, which became (also listed) IPHC in 2006 [https://50ans.in2p3.fr/timeline-iphc]. Thus CRN can be removed.

Scientists listed with affiliations to CEA/Gif-sur-Yvette, CNRS/Grenoble and IN2P3/Paris are actually affiliated to other institutes already present on the JINR list, to which CEA/CNRS or IN2P3 serve as funding organizations. Thus all three can be removed as duplicates.

Annecy-le-Vieux does not exist as a city since 2017, Laboratory of Annecy-le-Vieux for Particle Physics has been renamed shortly after that to Laboratory of Annecy for Particle Physics. It does not collaborate with JINR on the projects listed.

**Appendix B JINR-collaborator Institutes under International Sanctions as of 20/10/2024**

78 JINR-collaborator institutes based in Russia fall under sanctions: EU 19, USA 34, SW 19, UK 7, and Ukraine (UA) 70. The information below is taken from the OpenSanctions database:

1. Institute of Solid-State Physics of the Russian Academy of Sciences : EU, USA, Switzerland, UA
2. L.D. LANDAU INSTITUTE FOR THEORETICAL PHYSICS OF RUSSIAN ACADEMY OF SCIENCES: USA, Canada, Japan, UA
3. Joint Stock Company State Scientific Centre Research Institute of Atomic Reactor: Australia, Canada, UK, UA
4. Moscow Institute of Physics and Technology: EU, UK, USA, Canada, Switzerland, Japan, UA, New Zealand
5. NPP Istok: EU, USA, Canada, Switzerland, Japan, UA, New Zealand
6. NATIONAL RESEARCH CENTER KURCHATOV INSTITUTE: USA, UA
7. JOINT STOCK COMPANY SCIENTIFIC AND TECHNICAL COMPANY AZIMUT PHOTONICS: USA
8. Moscow State Technical University N.E. Bauman: USA, UA
9. Cryogenmash Public Joint Stock Company: USA, UA
10. A.M. PROKHOROV GENERAL PHYSICS INSTITUTE RUSSIAN ACADEMY OF SCIENCES: USA
11. FEDERAL STATE BUDGET SPACE RESEARCH INSTITUTE RUSSIAN ACADEMY OF SCIENCES / SRI RAS, USA
12. A.A. Kharkevich Institute for Information Transmission Problems (IITP), Russian Academy of Sciences (RAS): Canada, EU, USA, Switzerland, UA

13. BAIKOV INSTITUTE OF METALLURGY AND MATERIALS SCIENCE OF THE RUSSIAN ACADEMY OF SCIENCES: USA
14. PJSC "Institute of Electronic Control Machines named after. I.S. Bruk": USA, Japan, UA
15. Institute of Theoretical and Experimental Physics, EU, USA, Switzerland, UA
16. Keldysh Institute of Applied Mathematics of the Russian Academy of Sciences: EU, Switzerland, USA
17. FEDERAL STATE FINANCED INSTITUTION OF SCIENCE PHYSICAL HIGHER EDUCATION INSTITUTION NAMED AFTER P. N. LEBEDEVA OF THE RUSSIAN FEDERATION ACADEMY SCIENCES: EU, USA, Canada, Switzerland, Japan, UA
18. Moscow Aviation Institute: EU, USA, Switzerland, UA
19. NIKIET Joint Stock Company "N.A. Dollezhal Research and Development Institute of Power Engineering" EU, UK, USA, Canada, Switzerland, Australia, Japan, UA
20. Joint-Stock Company "Highly TechnologicalScientific-Research Institute of Non-OrganicMaterials named after AcademicianA.A. Bochvar": UK, UA, AU
21. SKOLKOVO INSTITUTE OF SCIENCE AND TECHNOLOGY: USA, Switzerland, Australia, UA, New Zealand
22. National Research University "MIET": UA
23. Federal State Budgetary Institution of Science "Institute of Archeology of the Russian Academy of Sciences": UA
24. JSC "Special Design Bureau of the Moscow Power Engineering Institute": USA, Canada
25. National Research University "Higher School of Economics": Canada, UA
26. A. O. Kovalevsky Institute of Biology of the Southern Seas of RAS: UA
27. Federal State Unitary Enterprise Dukhov Automatics Research Institute (VNIIA): EU, USA, UK, Canada, Switzerland, Australia, Japan, UA, New Zealand
28. Federal Technical Regulation and Metrology Agency: EU, USA, Switzerland, UA
29. FEDERAL RESEARCH CENTER INSTITUTE OF APPLIED PHYSICS OF THE RUSSIAN ACADEMY OF SCIENCES: EU, Swiss, USA, Canada, Japan, UA
30. Federal Research Center Boreskov Institute of Catalysis: EU, USA, Canada, Switzerland, UA, New Zealand
31. BUDKER INSTITUTE OF NUCLEAR PHYSICS OF SIBERIAN BRANCH RUSSIAN ACADEMY OF SCIENCES: USA
32. RZHANOV INSTITUTE OF SEMICONDUCTOR PHYSICS SIBERIAN BRANCH OF RUSSIAN ACADEMY OF SCIENCES: EU, USA, Canada, Switzerland, Japan, UA
33. State Scientific Center AO GNTs RF – FEI A.I. Leypunskiy Physico-Energy Institute: EU, USA, Switzerland, UA
34. Federal State Unitary Enterprise "Russian Federal Nuclear Center - All-Russian Research Institute of Experimental Physics": EU, UK, USA, Canada, Switzerland, New Zealand, Australia, UA
35. Federal State Unitary Enterprise "Russian Federal Nuclear Center - All-Russian Research Institute of Technical Physics named after Academician E.I. Zababakhin": EU, USA, UK, Switzerland, Australia, New Zealand, Japan, Canada, UA
36. FEDERAL STATE UNITARY ENTERPRISE CENTRAL RESEARCH INSTITUTE OF STRUCTURAL MATERIALS PROMETEY NAMED BY I.V. GORYNIN OF NATIONAL

RESEARCH CENTER KURCHATOV INSTITUTE: USA, Canada, New Zealand, Australia, UA
37. FEDERAL STATE FINANCED INSTITUTION OF SCIENCE PHYSICS AND TECHNOLOGY INSTITUTE NAMED AFTER A. F. IOFFE OF THE RUSSIAN FEDERATION ACADEMY OF SCIENCES: USA, Japan
38. Institute of High Energy Physics: EU, USA, Switzerland, UA
39. Federal State-Funded Educational Institution of Higher Education "Saint Petersburg Mining University": USA, UA
40. ISE SO RAN Institute of High-Current Electronics: EU, USA, Switzerland, UA
41. Private Institution "Iter-Center": UA
42. Federal State-Funded Educational Institution of Higher Education "Plekhanov Russian University of Economics": UA
43. Joint-Stock Company "State Specialized DesignInstitute": UA
44. Federal State Autonomous Educational Institution of Higher Education "Southern Federal University": UA
45. St. Petersburg State Electrotechnical University LETI named after VI Ulyanov (Lenin): UA
46. Northern (Arctic) Federal University: UA
47. Belgorod State National Research University : UA
48. South Ural State University: UA
49. Federal State Budget Educational Institution of Higher Education "Moscow State University named after M.V. Lomonosov": UA
50. Chechen State Pedagogical Institute: UA
51. Udmurt State University: UA
52. Baltic Federal University named after Immanuel Kant: UA
53. Pacific State University: UA
54. Kuban State University: UA
55. National Research University of Technology MISiS: UA
56. Peoples' Friendship University of Russia: UA
57. Nizhny Novgorod State University named after Lobachevsky: UA
58. Omsk State University named after Dostoevsky: UA
59. Perm State National Research University: UA
60. Samara National Research University named after Academician Korolyov: UA
61. Saratov National Research State University named after Chernyshevsky: UA
62. National Research University ITMO: UA
63. JointStock Company "Radium Institute namedafter V.G.Khlopin": UA
64. St. Petersburg Polytechnic University of Peter the Great: UA
65. St. Petersburg State University: UA
66. St. Petersburg State Forestry University named after S.M. Kirov: UA
67. Bashkir State University: UA
68. National Research Tomsk Polytechnic University: UA
69. National Research Tomsk State University: UA
70. Tula State University: UA
71. North Ossetian State University named after K.L. Khetagurov: UA

72. [Far Eastern Federal University](#): UA
73. [Voronezh State University](#), UA
74. [Northeastern Federal University named after M.K. Ammosov](#): UA
75. [Ural Federal University named after the first President of Russia Yeltsin](#): UA
76. [Volga State Technological University](#): UA
77. [Moscow State Pedagogical University](#): UA
78. [National Research Nuclear University "MEPhi"](#): UA